\documentclass[fleqn,10pt]{wlscirep}
\usepackage[utf8]{inputenc}
\usepackage[T1]{fontenc}
\usepackage{siunitx}
\usepackage{float}
\usepackage{amsmath}
\usepackage{amsfonts}
\usepackage{pdfpages}
\DeclareMathOperator{\sinc}{sinc}
\title{Investigating the Limits of Hard X-ray Coherence Length Measurement Employing Young's Double Slit Experiment.}

\author[1,2,*]{Rielly Castle}
\author[2]{Narayan Appathurai}
\author[1]{Nicholas Simonson}
\author[1]{Yasaman Sigari}
\author[1]{Mark Boland}
\author[2]{Feizhou He}
\author[2]{Chithra Karunakaran}
\author[2]{Jian Wang}
\author[2]{Beatriz D. Moreno}
\author[1,2,*]{Venkata S.C. Kuppili}
\affil[1]{University of Saskatchewan, Saskatoon, SK. S7N 5E2, Canada}
\affil[2]{Canadian Light Source, Saskatoon, SK. S7N 2V3, Canada}

\affil[*]{rfc020@usask.ca}
\affil[*]{charan.kuppili@lightsource.ca}

\begin{abstract}
Young's double slit experiment has been the most explored technique to gauge the coherence properties of a given system. The limits of this technique in characterizing spatial coherence properties of high emittance, hard x-ray synchrotron sources have been performed at the BXDS-IVU beamline, Canadian Light Source (CLS). High emittance synchrotron sources have always been assumed to possess sub-optimal coherence properties, especially in the hard X-ray regime. While this is largely true, it is very important to understand the limits of coherence for these existing high emittance sources. We have demonstrated that the Young's double slit experiment has harsher limits than what is normally expected in the form of inherent ambiguity. We present data obtained at multiple energies in both the horizontal and vertical direction leading to a thorough understanding of the fundamental limitations of employing Young's double slit experiment to characterize inherently low coherence length systems. We have put forth a novel numerical technique to estimate the source size directly from the Young's double slit interference patterns. With these results, we have demonstrated that CLS has functional coherent beam properties in the hard X-ray regime with Spatial coherence lengths ranging from 5.37~\si{\micro\meter} to 17.61~\si{\micro\meter} in the horizontal direction. The same in the vertical direction was at least 3 times larger. Finally, we present theoretical calculations showcasing the limits of Young's double slit experiment in characterizing diffraction limited sources. 

\end{abstract}
\begin{document}

\flushbottom
\maketitle
%
%
\thispagestyle{empty}

\section*{Introduction}
\label{section:Sec1}
In recent years, multiple 3$^{\text{rd}}$ and 4$^{\text{th}}$ generation synchrotron sources have been built to further advance fields like materials science, biology, and chemistry~\cite{Horiuchi2015, Martensson2014}. Due to the intrinsic coherence properties of these light sources, the use of coherence-based techniques such as coherent diffraction imaging (CDI), X-ray holography, Bragg CDI, and ptychography have become a mainstay~\cite{Lo2018, Soltau2021, Hitchcock2015, Thibault2008, Miao1999, Kuppili2017}. Consequently, it is of great interest to characterize the coherence properties of these light sources. \newline

\begin{figure}[h!]
    \centering
    \includegraphics[width=15cm]{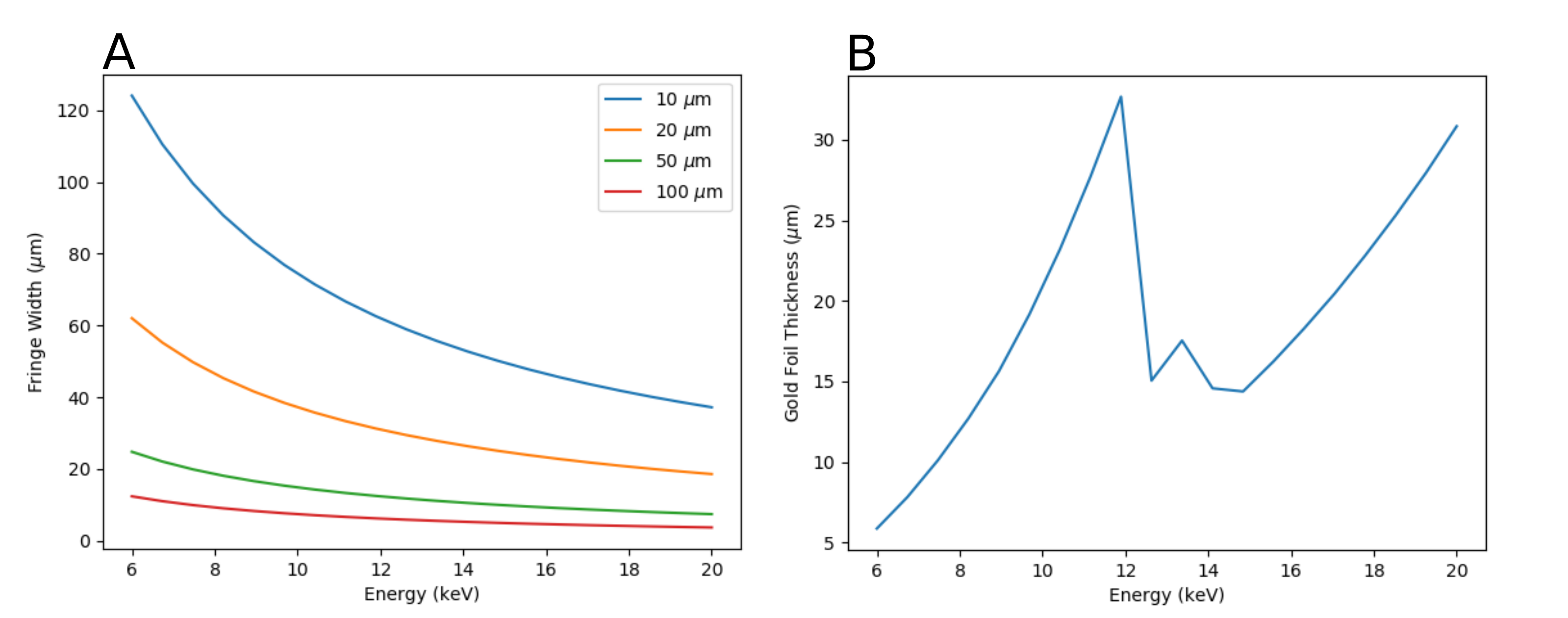}
    \caption{A shows the relationship between fringe width ($\mu$m) obtained in a YDS experiment as a function of energy for various slit separations (10,20,50,100 $\mu$m), when the detector is positioned 6 meters from the slits. The fringe width reduces with increase in energy as well as increase in slit separation thus imposing constraints on sampling the fringe pattern. B shows the necessary thickness of gold foil that is needed to be employed in order to obtain 1\% transmission as a function of energy. Except for an interim gold absorption edge, the thickness increases with an increase in energy.}
    \label{fig:clng_yds}
\end{figure}

\noindent
Young's double slit (\textbf{YDS}) experiment is probably the most employed experiment to measure coherence properties in a given system ~\cite{Leitenberger2001,Vartanyants2010,DOHYUNG2019,Takayama1998a}.  The apparatus involves positioning a completely opaque object containing two openings or \textit{slits} of given width which are themselves separated by a certain distance. The coherence length is typically measured by analyzing the visibility of interference patterns formed in a YDS apparatus~\cite{Leitenberger2001}. Although this seems straightforward with regard to visible light or even soft-X-rays, the same in the hard-X-ray regime poses significant challenges. Firstly, thin metal foils of large atomic number are capable of blocking the majority of soft X-rays for thicknesses of even a few microns. The same in the hard X-ray regime would range in tens or even hundreds of microns (Fig. \ref{fig:clng_yds}). What complicates the issue of thickness further would be the necessity to fabricate precise slits that are a few microns wide into these foils. There is usually a limit on the aspect ratio with which such slits can be fabricated into metal foils. The state-of-the-art, cutting edge fabrication techniques can still have a limit of 10-20 (foil thickness/slit width). The fundamental reason behind the need to employ micron-scale slit widths is embedded in the relationship between the wavelength of radiation interacting with the slits, slit width, slit separation and the observed interference fringes in the detector plane \cite{optics_book}. The larger the width of the slits, the smaller the fringe width on the detector plane. Owing to the physical limitations of pixel sizes in currently available detectors, the slit widths and the separation between the slits need to be in the micron range for the pattern to have optimal sampling in the detector plane. It is also important to note that the slit separation plays a fundamental role in its ability to measure larger coherence lengths, the larger the coherence length one would like to measure, the larger the slit separation one needs to employ. This will further have an effect on fringe widths that constitute the interference patterns wherein the fringe widths get smaller with an increase in the slit separation, further imposing constraints on the pixel sizes of employed detectors (Fig \ref{fig:clng_yds}). Pixel sizes in state-of-the-art imaging detectors usually deployed in tomography beamlines can get as small as 250 nm \cite{bmitcam,optiquepeter}. While double slit interferometry is considered as the most reliable and well-established technique to measure coherence properties, other techniques have been introduced to overcome various challenges of measuring coherence length in the hard X-ray regime~\cite{Jacques2012, Marathe2014,Kohn2000}.\\

\noindent
The emittance of a synchrotron source is defined to be the area occupied by particles within the beam in position-momentum phase space. This quantity is related to the coherence properties of the generated X-ray beams in the sense that they will be smaller and more coherent in the case of a low emittance source~\cite{emittance_book}. As such, it has been typically assumed that 3$^{\text{rd}}$ generation synchrotron sources with low emittance are better suited to exploit coherence properties, especially in the hard X-ray regime. At this juncture, it is also important to mention long beamlines~\cite{Pesic2013}. By utilizing the van Cittert–Zernike theorem, these beamlines achieve long spatial coherence lengths by simply moving away from the undulator source. High-emittance sources such as the Canadian Light Source (CLS) have largely been neglected from these types of studies. Generally, high-emittance synchrotron sources are assumed to generate X-ray beams with impractically low coherence lengths. To some extent, one can still increase spatial coherence length by temporarily making the source smaller, for example, by employing slits that spatially constrain the source extent. This is not very different from what long beamlines do by moving away from the source or what diffraction-limited sources do by fundamentally reducing the size of the source\cite{Pesic2013,Eriksson2014}. However, the long beamlines and diffraction-limited sources do this at a much grander scale, achieving impressive coherence lengths even in the hard X-ray regime.\\ 
\newline
In this study, we investigate the limits of employing the YDS experiment to measure spatial coherence properties in the hard X-ray regime. We will demonstrate that YDS has inherent ambiguity beyond a certain value of visibility, and we will also propose the necessary protocols needed to overcome these limitations. We will back this up with multi-energy, bi-directional spatial coherence measurements at the BXDS-IVU hard X-ray beamline, Canadian Light Source. In the process, we will also demonstrate that high-emittance synchrotron sources still have functional coherence lengths. Finally, we will show that, as expected, one can improve coherence length by constraining the source extent. 



\section*{Theory}
\label{section:Sec2}

\subsection*{Young's Double Slit Experiment and Spatial Coherence Length}
\label{section:Sec2.1}

In practice, the coherence properties are determined using the fringe visibility of the intensity distribution we observe during a Young’s double slit experiment. The interferometric fringe visibility can be expressed mathematically using the function~\cite{optics_book, Jackson2018, optics_book2}, 
\begin{equation}\label{eq:1}
    V = \frac{I_{max} - I_{min}}{I_{max} + I_{min}} = \left|\sinc\left(\frac{\pi Dd}{\lambda R}\right)\right|,
\end{equation}
where $I_{\text{max}}$ and $I_{\text{min}}$ are the maximum and minimum fringe intensities, respectively. $D$ is the source size, $d$ is the double slit separation, R is the source-to-slit distance, and $\lambda$ is the X-ray wavelength. The function $\sinc(x)$ is a special function that is defined as
$\sin(x)/x$. The intensities can be expressed with the equations,
\begin{equation}
    I_{max} = I_{1} + I_{2} + 2|\gamma |\sqrt{I_{1}I_{2}},
\end{equation}
\begin{equation}
    I_{min} = I_{1} + I_{2} - 2|\gamma |\sqrt{I_{1}I_{2}},
\end{equation}
where $I_{1}$ and $I_{2}$ are the observed intensities from the first and second slit. The quantity $\gamma$ is known as the degree of coherence. We see that when $\gamma =1$, the sources are fully coherent, completely incoherent when $\gamma =0$, and partially coherent sources take on some value in between. Under the assumption that the wavefront from both of the slits is monochromatic and that each wavefront holds the same polarization, we can express the visibility function as,
\begin{equation}
    V = \frac{2|\gamma |\sqrt{I_{1}I_{2}}}{I_{1} + I_{2}}.
\end{equation}

\noindent
Spatial coherence length is \textit{typically} defined to be the slit separation required to reduce the fringe visibility to 0 for a given source size. Alternatively, it can also be the source extension required to reduce the fringe visibility to 0 for a given slit separation. Using the relationship between coherence length and the size of the source extension, spatial coherence length can be calculated with the equation~\cite{Xray_phys_book}
\begin{equation}\label{eq:coh_len}
    l_{c} = \frac{\lambda R}{2D}.
\end{equation}

\subsection*{Synchrotron Sources as Gaussian–Schell Model Sources}
\label{section:Sec2.2}
Synchrotron sources can be approximated to Gaussian-Schell model (\textbf{GSM}) sources where the source can be thought of as an extended, stochastic elliptical disk of uniform density~\cite{derVeen2004, Nugent2010, Kuppili2019}. Each point contained within this disk behaves like a fully coherent point source capable of emitting X-rays. The point sources are, however, assumed to be mutually incoherent to each other. The center of this source and the center of the YDS are assumed to be on the same axis. A plane wave $q_{0}(X,Y)$ emitted from the central point source reaches the YDS with the normal of the wavefront parallel to the normal of the YDS. A wave emitted from a point $(D_X,D_Y)$ on this source reaches the YDS with its normal at an angle $\theta$ with respect to the normal of YDS. The intensity of the far-field diffraction pattern of such an interaction will be a shifted double slit intensity function. Without loss of generality, one can safely neglect the effect of vertical shifts while considering horizontal slits and vice versa. As each of the point sources is assumed to be mutually incoherent, the resultant intensity of such a GSM source interacting with YDS amounts to the summation of shifted intensities of double slit intensity functions. By forward modeling the resultant intensity of far-field diffraction patterns and cross-checking them with experimentally observed intensity patterns, one obtains the maximum shifted distance in the detector space, $\eta$. Numerically, this corresponds to summing up the double slit intensity functions shifted along the detector space. $\eta$ is given by $L\theta$ where $L$ is the slit-to-detector distance. Making a small angle approximation, one obtains the relationship between GSM source size (D) and $\eta$ as $\eta=D\frac{L}{R}$. One can then obtain the information on coherence length through equation \ref{eq:coh_len}. More details regarding data analysis will be discussed in Section 3.2.

\section*{Results}
\label{section:Sec4}

\subsection*{Fringe Visibility and Source Size Calculations.}
\label{section:Sec4.1}
\begin{figure}[h!]
    \centering
    \includegraphics[width=13cm]{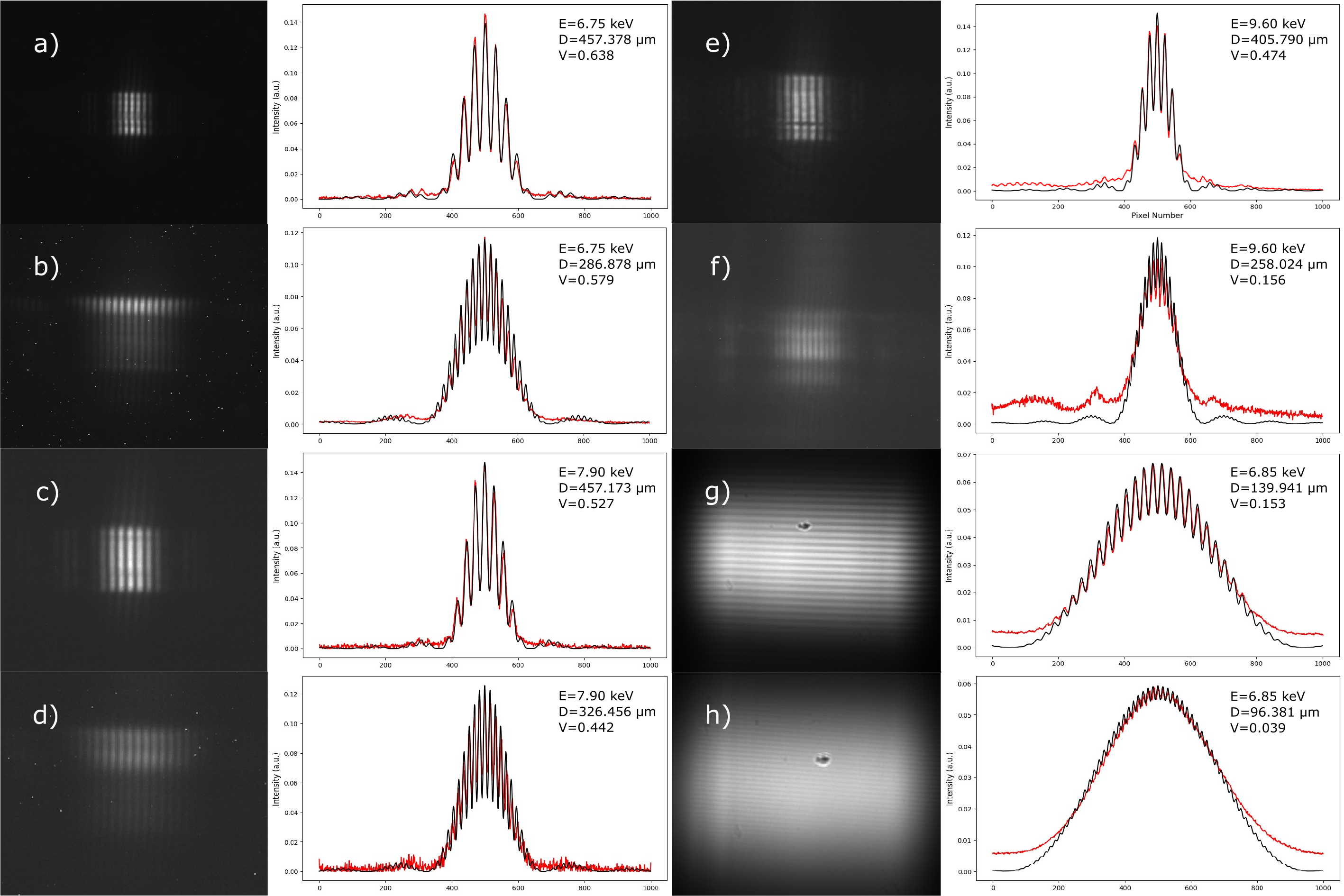}
    \caption{The observed interference fringes for each energy and double slit used during this experiment. In all cases, the red curve shows the experimental data, while the black curve is the numerically fitted interference fringes. Subfigures (a-b) show the interference fringes at E=6.75~ \si{keV} and slit separations of 10~\si{\micro \meter} and 20~\si{\micro \meter}. Subfigures (c-d) show the interference fringes at E=7.9~\si{keV} and slit separations of 10~\si{\micro \meter} and 20~\si{\micro \meter}. Subfigures (e-f) show the interference fringes at E=9.6~\si{keV} and slit separations of 10~\si{\micro \meter} and 20~\si{\micro \meter}, all the patterns shown to this point correspond to horizontal coherence length measurements. Subfigures (g-h) show interference fringes from the vertical double slit measurements taken at $E=6.85$~\si{keV}. The slit separations are 50~\si{\micro \meter} and 100~\si{\micro \meter}, respectively. The x-axis in all plots represents pixel numbers. The quantity annotated in each figure is the energy used for the measurement and the source size calculated by the fitting procedure using goodness of fit as a figure of merit to determine the source size. Subfigures (a-f) represent fringes observed during horizontal spatial coherence measurements, while subfigures (g-h) represent the fringes observed during the vertical spatial coherence length measurement.}
    \label{fig:AllFringes}
\end{figure}

\begin{figure}[h!]
    \centering
    \includegraphics[width=13cm]{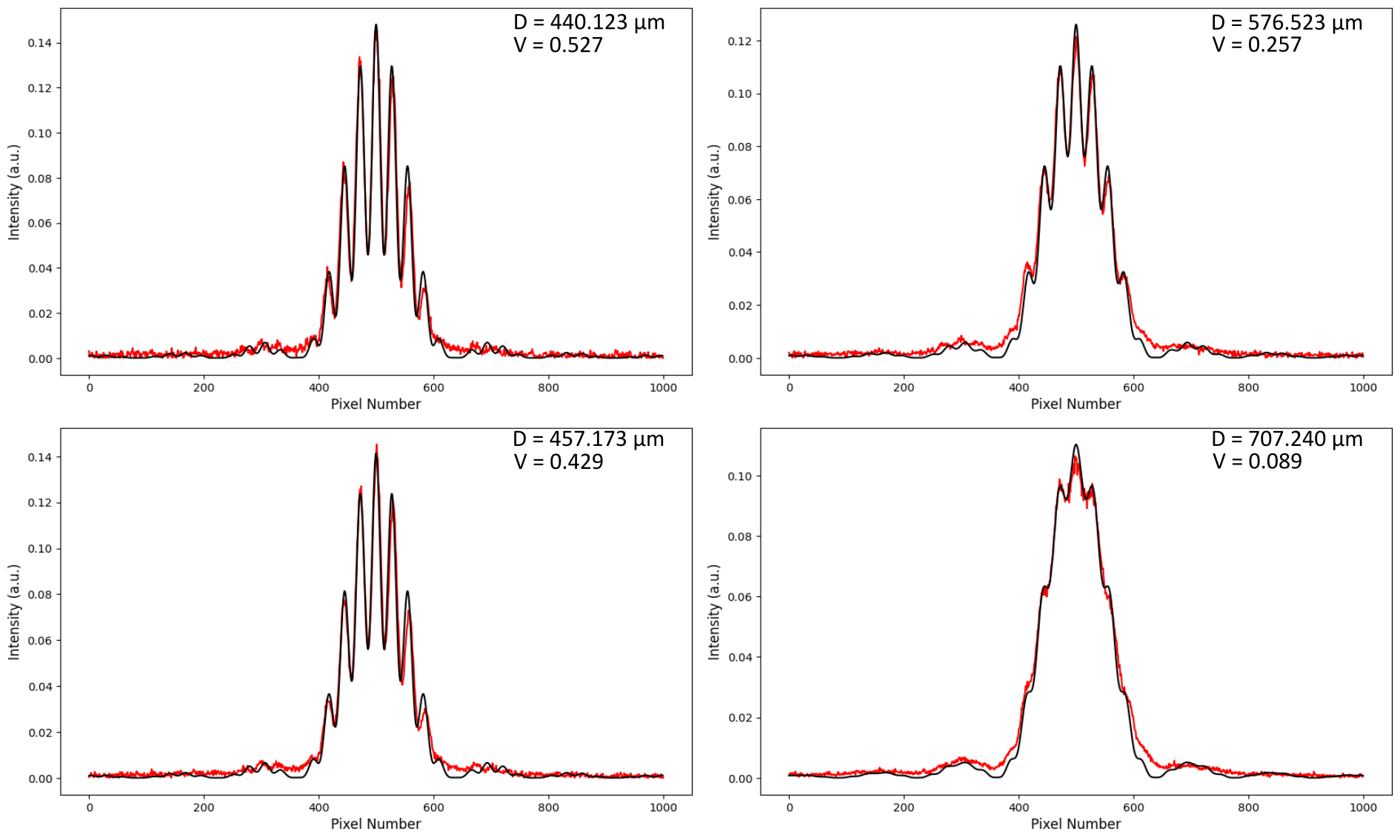}
    \caption{Displayed are the interference fringes made at $E=$\SI{7.9}{keV} using \SI{10}{\micro \meter} double slits as the source size is increased. In all cases, the red curve represents the observed intensity distribution, while the black curve represents the numerically fitted curve. All fringes shown are from horizontal spatial coherence measurements.}
    \label{fig:FringeBlurring}
\end{figure}

Interferometric fringes have been observed at all the energies explored during the experiment. Coherence length calculations were not performed on data taken at energies above $E=15$~\si{keV} due to sampling issues and the transmission of X-rays through the double slit array. The observed intensity distribution of the interference fringes using $E=20~\si{keV}$ is shown in the Supplementary Material (Fig. S5). This demonstrates the dual limitations of foil thickness, and fringe sampling at higher energies, as illustrated in Fig.~\ref{fig:clng_yds}. In accordance with the van-Cittert Zernike theorem, as expected, the observed fringe visibility decreased as the source size increased. Fringe visibility also decreased with an increase in energy. Fig.~\ref{fig:AllFringes} shows 2D profiles of observed fringes and their fitted intensity profiles for each energy and double slit. Fig.~\ref{fig:FringeBlurring} shows an example of the blurring of fringes as the source size is increased.\\

\noindent
Source extension sizes obtained by both the goodness of fit method (\textbf{GOF}) and the interpolation method (\textbf{INTPO}) strongly agree with the theoretical model (check methodology section for more information). Both methods produced near-identical results, differing only by a few microns in terms of calculated source sizes. The divergence of values between the methods increased with decrease in visibility (Supplementary Material, Fig S13). We suspect that this is a result of a lack of sampling near the first minimum region for a given visibility vs source size curve, and as one can expect, interpolation is sub-optimal under such circumstances (Fig.~\ref{fig:vis_sos_intpo}).\\

\noindent
The measurements in the vertical direction confirmed the expected emittance, source sizes in the vertical direction, and as expected, they were at least 3 times smaller when compared to the horizontal direction. One can also see the need for larger slit separations in order to measure small source sizes, while the curves pertaining to smaller slit separations taper off at relatively larger source sizes the same is comparatively sharper with an increase in slit separation.\\

\noindent
We observed a mild asymmetry in the experimental far-field diffraction patterns collected at low energies, which increased at higher energies. SEM imaging has revealed that the slit widths are not identical in the double slits. Detailed results of the physical double slit measurements are provided in the Supplementary Material (Section S2). Measurements indicate that the slit widths are $3.27\pm 0.02$~\si{\micro\meter} and $2.59\pm 0.02$~\si{\micro\meter}. Consequently, the asymmetry can be attributed to higher intensity passing through the wider slit (Supplementary Material, Section S4).\\

\subsection*{Additional Limits Imposed by Inherent Ambiguity of YDS Experiment.}
\label{section:Sec4.2}

\begin{figure}[h!]
    \centering
    \includegraphics[width=15cm]{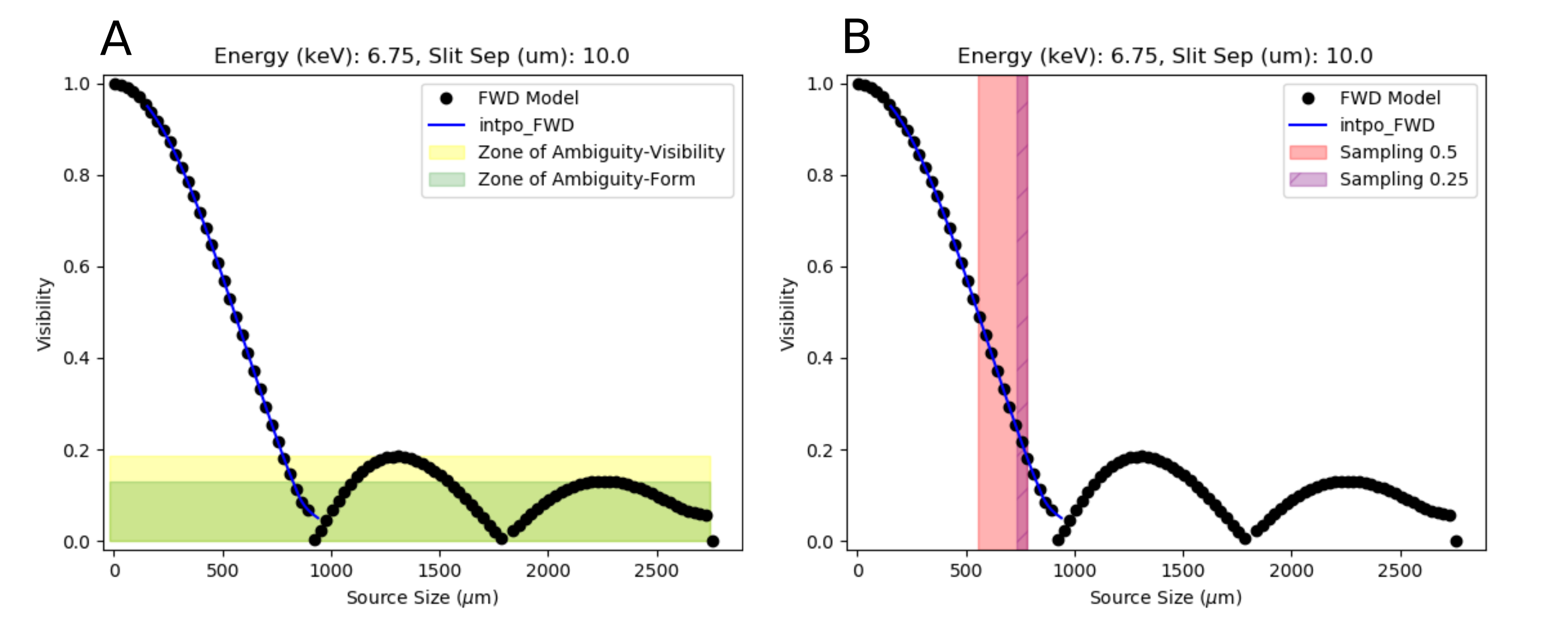}
    \caption{A shows the concept of Zone of Ambiguity (yellow, green region) in a given Visibility vs Source size curve, here forward model is denoted by black dots (.), interpolated curve is denoted by continuous blue line. Below a certain visibility value (maximum visibility before ambiguity), the source size is fundamentally ambiguous, and the actual source size can't be ascertained accurately. Further to this, the ambiguity itself can be classified into two classes: ambiguity by visibility values and ambiguity by form.
    One obtains identical visibility values at multiple source sizes under the zone of ambiguity-visibility, but the form itself is slightly different among various lobes. The forms are also identical between some lobes under the zone of ambiguity-form (supplementary figures S11, S12). The zones shown in this figure are with respect to lobe 1 and it is important to understand that within lobe 2-3 (same in 4-5 and 6-7), the interference patterns are ambiguous by both visibility and form. It should be noted that this does not strictly correspond to a lack of coherence itself. B shows sampling requirements during the experiment in order to obtain accurate visibility vs source size curves. Here, the sampling requirement from an initial 0.5 visibility is denoted by the pink region while the sampling requirement from 0.25 visibility is denoted by the purple region. The forward model and interpolated curves are denoted the same as A.}
    \label{fig:zoa_samp}
\end{figure}

It is typically assumed that the source is fully incoherent once the visibility of fringes reduce to zero, however, this is not necessarily the case and is, strictly speaking, contextual. While visibility might be zero for a given pair of slits, the source can still be small enough to exhibit high visibility values for smaller slit widths. One should therefore not conclude the lack of coherence properties based on measurements pertaining to a single pair of slits alone. YDS imposes additional limits on source size measurements owing to the inherently ambiguous nature of the visibility vs source size curve under a certain visibility value (Fig \ref{fig:zoa_samp}). This value, which we define as maximum visibility before ambiguity (\textbf{MVBA}), can be as large as 0.2, below which the estimations enter a zone of ambiguity where source sizes cannot be determined with certainty. Its source size counterpart that corresponds to MVBA is defined as minimum source size before ambiguity (\textbf{MSBA}). Our calculations further show that there are two kinds of ambiguity (supplementary figures S11, S12), namely ambiguity by visibility (\textbf{MVBA-Visibility}) and ambiguity by form (\textbf{MVBA-Form}) and their source size counterparts. In the case of ambiguity by visibility, very different source sizes give identical visibility values, and unassuming operators can make false estimations between widely different source sizes going purely by visibility value. The source sizes can still be told apart from the form of the central peak of the pattern (supplementary figures S11, S12, \cite{Jackson2018}), but it is just as likely to mistake this for a camera artifact.\\

\begin{figure}[h!]
    \centering
    \includegraphics[width=15cm]{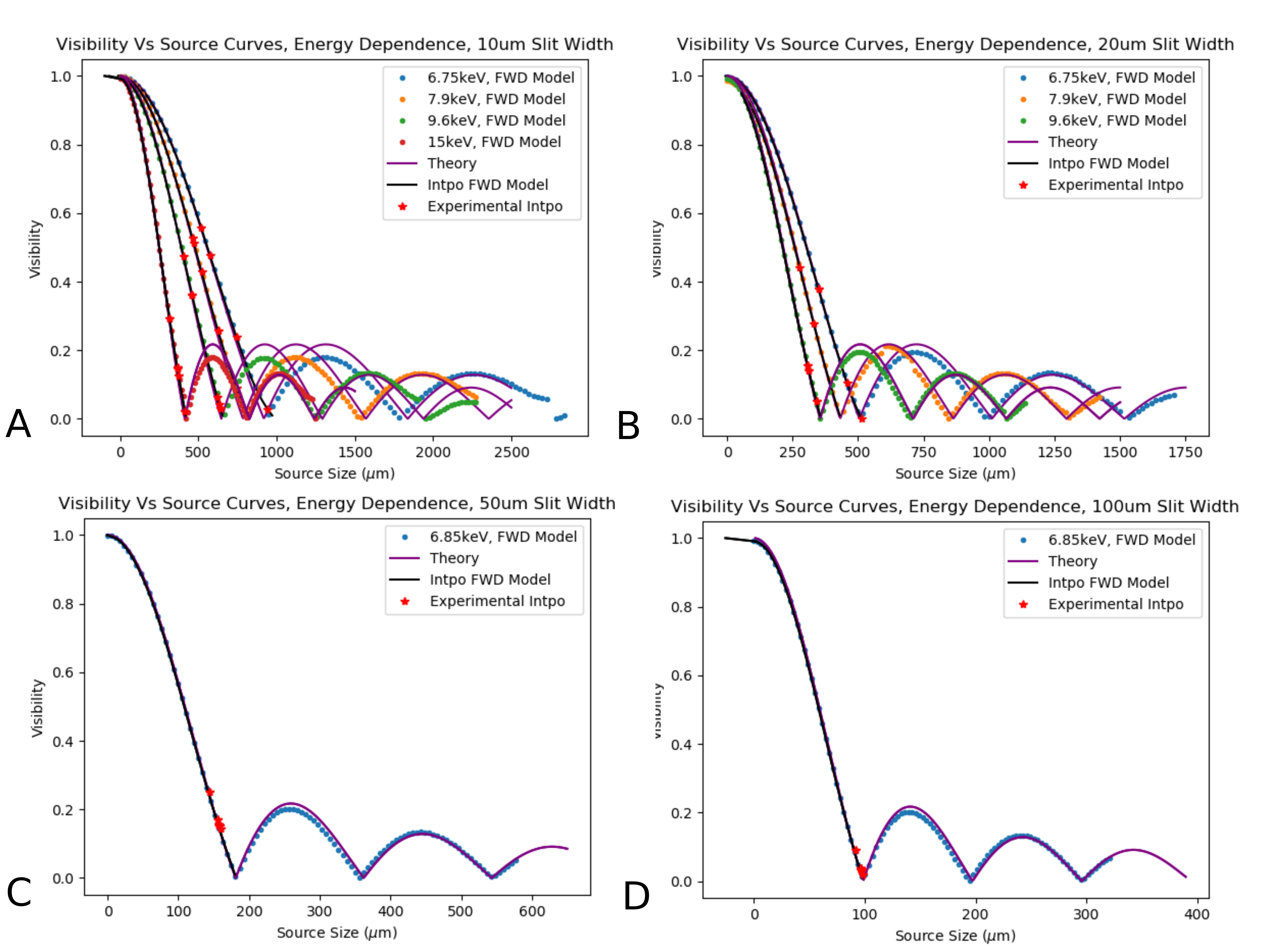}
    \caption{Figure showing Visibility vs Source Size behavior obtained from the proposed forward modeling approach, A,B,C,D shows various curves classified by slit separation. Forward model curves for each slit separation are further compartmentalized by energy (keV) parameter. While the forward model is denoted by \textit{colored} dots (.), the interpolated curves for each of these curves are denoted by a continuous black line throughout the figure. Theoretically expected curves generated using equation \ref{eq:1} are denoted by a continuous purple line throughout the figure. The experimental data points are denoted by red stars (*) on top of the interpolated segments of respective energy, slit width throughout the figure. The data shown in C and D pertain to coherence length measurements taken along the vertical direction.}
    \label{fig:vis_sos_intpo}
\end{figure}

\noindent
In order to better understand this phenomenon, we divide the visibility vs source size curve into seven lobes. Lobe 1 is the region of high visibility and low source sizes, however, as one descends into lower visibility states, the visibility values start matching up to the low visibility states of lobe 2-7. As one enters the zone of ambiguity by visibility (yellow region as shown in Fig \ref{fig:zoa_samp} A), for the same visibility value, there are at least 2 source sizes. At this juncture, it is important to make a distinction between ambiguity by visibility and ambiguity by form. Between lobe 1 and lobe 2,3 though the visibility values concur, one will still be able to tell things apart owing to the form of the central peak of the interference pattern. It should however be noted that the form also matches between lobe 1 and lobe 4,5 (supplementary figures S11, S12). Along with this, within the lobes 2-3 (also 4-5, 6-7), there is an ambiguity between different source sizes for the same visibility and form. It is, therefore, important to consider the following protocols when employing YDS for coherence measurements. One should strive to obtain visibilities above the MVBA value for a given slit separation and energy, higher the value better. In the scenario where the value can't be attained it will be best to move on to smaller slit separation or energy. While it would still be possible to make some assessment for visibility values between MVBA-Visibility and MVBA-Form owing to the difference in the form of the central peak, beyond this, the source size values will not be able to be ascertained accurately. Ambiguity plays a crucial role, especially in characterizing systems with low coherence properties. We took extreme care in calculating all the scenarios beforehand in order to arrive at an accurate picture with coherence length measurements. Between obtaining a visibility value below MVBA for a larger slit separation and a visibility value above MVBA for a smaller slit separation the later option would be better as in the former case there is inherent ambiguity while both the measurements might have come from the same source size. When one looks back at the collected data (Fig \ref{fig:vis_sos_intpo}) with fresh perspective, one can see that the maximum visibility values for a given energy and slit separation were kept above the MVBA. In cases where this couldn't be achieved (Fig \ref{fig:vis_sos_intpo}, 20$\mu$m slit separation, 9.6 keV) we consolidated the source sizes by looking into visibility values at lower slit separations (Fig \ref{fig:vis_sos_intpo}, 10$\mu$m slit separation, 9.6 keV).\\    

\begin{figure}[h!]
    \centering
    \includegraphics[width=15cm]{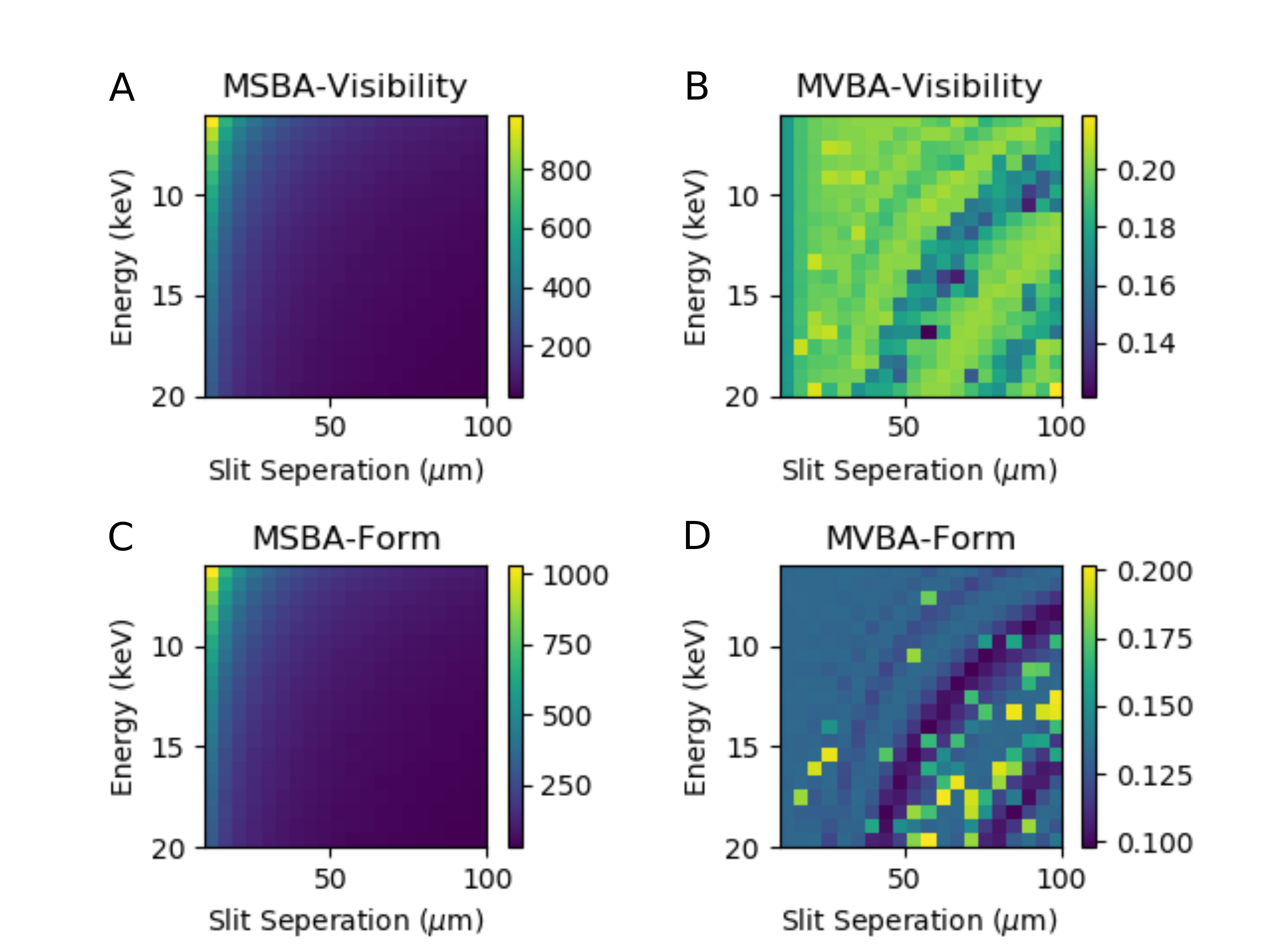}
    \caption{A,C shows dependence of minimum source size before ambiguity by visibility values and form respectively (\textbf{MSBA-Visibility, MSBA-Form}) as a function of energy and slit separation. Both of these quantities reduces with increase in energy as well as slit separation. The color bar in A,C shows source size in $\mu$m. B,D shows dependence of maximum visibility before ambiguity by visibility values and form respectively (\textbf{MVBA-Visibility, MVBA-Form}) as a function of energy and slit separation. The values vary around the value of 0.2 in the case of MVBA-Visibility, the same for MVBA-Form varies around 0.13 across the board. The color bar in B,D denotes visibility values.}
    \label{fig:amb_cur}
\end{figure}

\noindent
We have carried out a large-scale analysis of ambiguity and its effect as a function of energy and slit separation (Fig \ref{fig:amb_cur}). We observe that the MVBA-Visibility value stays constant varying slightly around .18, the same with MVBA-Form is .13 we see small variations owing numerical errors. The same with MSBA-Visibility, MSBA-Form give source sizes below which estimations can't be made accurately. It is important to note that MSBA value decreases with increase in energy as well as slit separation. One should therefore traverse along the decreasing energy, slit width separation values until one moves above the MSBA values thus reaching values where the source sizes can be estimated without ambiguity.        

\subsection*{Coherence vs Intensity, Effect of Sampling.}
\label{section:Sec4.3}
\begin{figure}[h!]
    \centering
    \includegraphics[width=15cm]{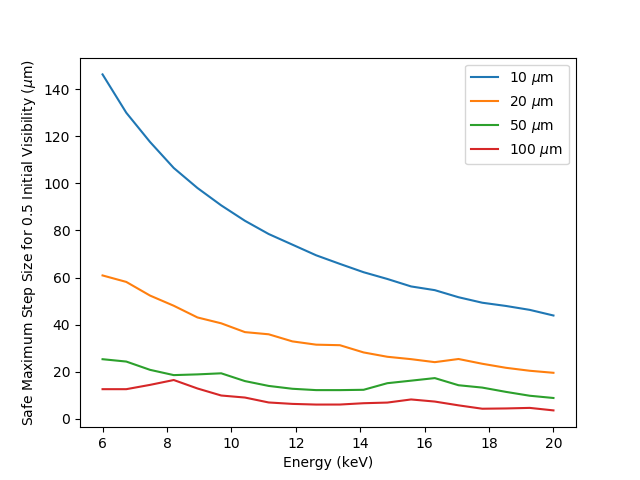}
    \caption{Figure shows sampling requirement from 0.5 initial visibility as a function of energy and slit separation. We can see that the sampling requirement get stricter with increase in energy and slit separation. It is also imperative that the requirement is stricter with regards to slit separation when compared to energy.   }
    \label{fig:samp_cur}
\end{figure}

It is a common practice to optimize the intensity of X-rays while compromising on coherence properties in a well-known, controlled manner. One normally adjusts the equivalent of front end slits (white beam slits in our case Fig \ref{fig:eFig1}) to get an optimal combination (Fig \ref{fig:FringeBlurring}). However, as one can expect, the ambiguity poses further challenges regarding this process. If the adjustment step is large enough that the visibility enters the zone of ambiguity, one will again face the issue of ambiguity in finding out the adjustment step size or sampling. The same is demonstrated in Fig.~\ref{fig:zoa_samp} B, the step sizes depending on the initial state of observed visibility should be smaller than the difference between MSBA and the source size corresponding to initial visibility. As the figure demonstrates, the smaller the existing visibility value, the smaller the sampling should be in order to stay above the MVBA, thus avoiding the ambiguity of step size. The best case scenario, therefore, would be to start the highest possible visibility state and take small steps to stay above MVBA as much as possible. This will also allow one to test the accuracy and repeatability of one's slit system after which one can be confident regarding intensity to coherence optimization process. Studying the same over large scale to gauge its behavior as a function of energy and slit sizes show (Fig \ref{fig:samp_cur}) that the step size gets smaller with increase in energy as well as with increase in slit separation distance. One can also observe that the step size reduces in a more drastic manner along the slit separation front when compared to energy. It is, therefore, advisable to carry out the testing process at lower energies. Again, looking back into the experimental data with this additional constraint (Fig \ref{fig:vis_sos_intpo}) best case scenario was observed at 10$\mu$m separation, low energies where both the visibility and sampling were optimal. As explained previously, 20$\mu$m slit separation at 9.6 keV and 50$\mu$m slit separation, 6.75 keV was still acceptable in terms of obtaining source size values without ambiguity. The worst case scenarios were 20$\mu$m slit separation at 9.6 keV and 100$\mu$m slit separation, 6.75 keV where both the visibility and sampling was subpar. If not for the data with lower slit separation, it would have been extremely hard to accurately determine the source size, coherence length properties.       




\subsection*{Coherence Length Profile of BXDS-IVU Beamline}
\label{section:Sec4.4}
\begin{figure}[h!]
    \centering
    \includegraphics[width=13cm]{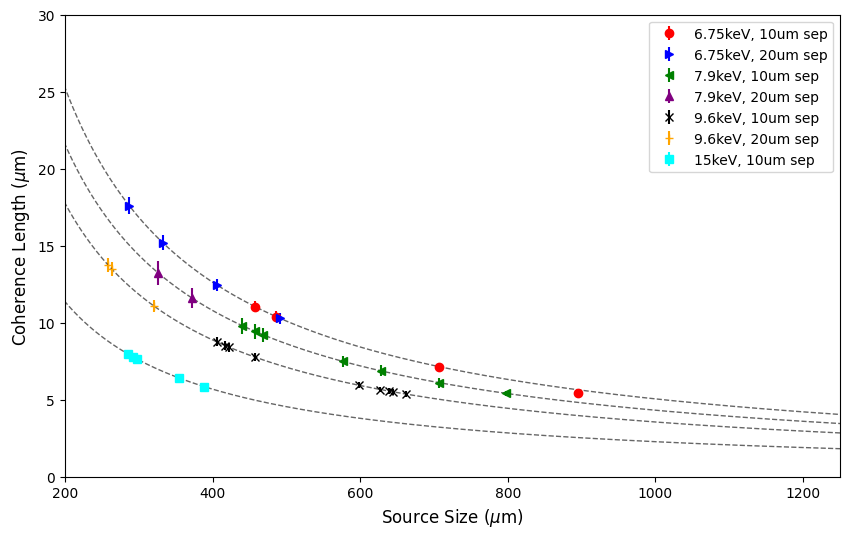}
    \caption{The measured horizontal spatial coherence length as a function of source size for all target energies. The dashed lines indicate the expected theoretical values. The error bars have been enlarged by a factor of 15 for better visibility.}
    \label{fig:CohLen}
\end{figure}

The observed fringe visibility ranges from 0.008 at its lowest to 0.56 at its highest. The spatial coherence length in the horizontal direction ranges from 5.37~\si{\micro m} to 17.61~\si{\micro m}, where the largest values are observed at lower energies using the 20 \si{\micro m} double slit. The calculated coherence lengths for each energy are shown in Fig.~\ref{fig:CohLen} as a function of source size. Additionally, the theoretical curves for spatial coherence length are plotted for comparison. The measured spatial coherence lengths are in excellent agreement with theoretical values. The same in the vertical direction was 3 times larger (Supplementary Material, Fig S6).

\section*{Discussion}
\label{section:Sec5}
\subsection*{Towards Coherence-Based Imaging.}
\label{section:Sec5.1}
Quantitative measurements of coherence length have been carried out in the hard X-ray regime. This led to a thorough characterization of the coherence properties of the CLS BXDS-IVU beamline in both horizontal and vertical directions. The success of this experiment indicates that the CLS has functional coherent beam properties in the hard X-ray regime. By extension, similar high-emittance 3$^{\text{rd}}$ generation synchrotron sources should likely be able to provide comparable coherence properties. The results shown in this study can be further improved with machine upgrades. Specifically, reducing the size of the electron beam should show significant improvement in the coherence properties of the generated X-rays. A direct application of these results is to begin to explore diffractive imaging techniques in the hard X-ray regime. Of particular interest would be establishing the capability of hard X-ray ptychography and ptycho-tomography\cite{Dierolf2010,Guizar-Sicairos2011a,Kuppili2024}. If successful, this achievement would be invaluable to scientists in a wide range of fields.

\subsection*{Limitations of Young's Double Slit Experiments and way Forward.}
\label{section:Sec5.2}
Young's double slit experiment (\textbf{YDS}) has immense potential to measure coherence properties of any given system. We have focused on various intricacies of characterizing low coherence systems employing YDS, its limitations and various protocols to mitigate them and obtain accurate estimates of source sizes of such systems. However, it should be noted that our protocols are inherently \textit{operations} intensive; the experiments demonstrated in this experiment needed upwards of 15 shifts (shift = 8 hrs). We think it will be extremely beneficial and productive if coherence length over multiple slit separations can be condensed into a single step. The same can be said about the lack of multi-dimensionality of our setup; an experiment that condenses multi-slit, multi-dimensional information into a \textit{single frame} will be much desired. We recognize that our existing forward model will have to be updated, upgraded so as to catch up with such a versatile experiment. In this context, we will take the opportunity to recognize the elegance and versatility of near-field techniques capable of characterizing coherence lengths along multiple spatial directions at one go \cite{Marathe2014}. Near-field techniques promise many advantages over far-field techniques like YDS. Firstly, the fringe visibility based detector sampling constraints are much more relaxed compared to YDS. As the near-field techniques do not need strictly opaque substrates, fabrication limitations are comparatively relaxed\cite{Marathe2014,Kohn2000}. It is however important to note that far field techniques are easier to forward model given that they can be reduced to a Fourier transform of sample substrate (double slits) while the same with near field is not so straight forward \cite{derVeen2004}. It has been suggested that for a single slit, the far-field and near-field patterns can be derived from each other through corrections \cite{Jacques2012}. We think there is a need to put some of these existing near-field, far-field coherence length measurement techniques on a scale so as to evaluate the pros and cons of each of these techniques. We have largely limited ourselves in exploring an energy range of 6-20 keV. Strictly speaking, we faced issues characterizing coherence properties at 20 keV owing to fabrication limitations. This will only get worse at higher energies, one should start thinking of innovative methodologies to characterize coherence lengths at ultra-high energies (>35 keV).     


\subsection*{Young's Double Slit Experiment and Diffraction Limited Sources}
\label{section:Sec5.3}

\begin{figure}[h!]
    \centering
    \includegraphics[width=13cm]{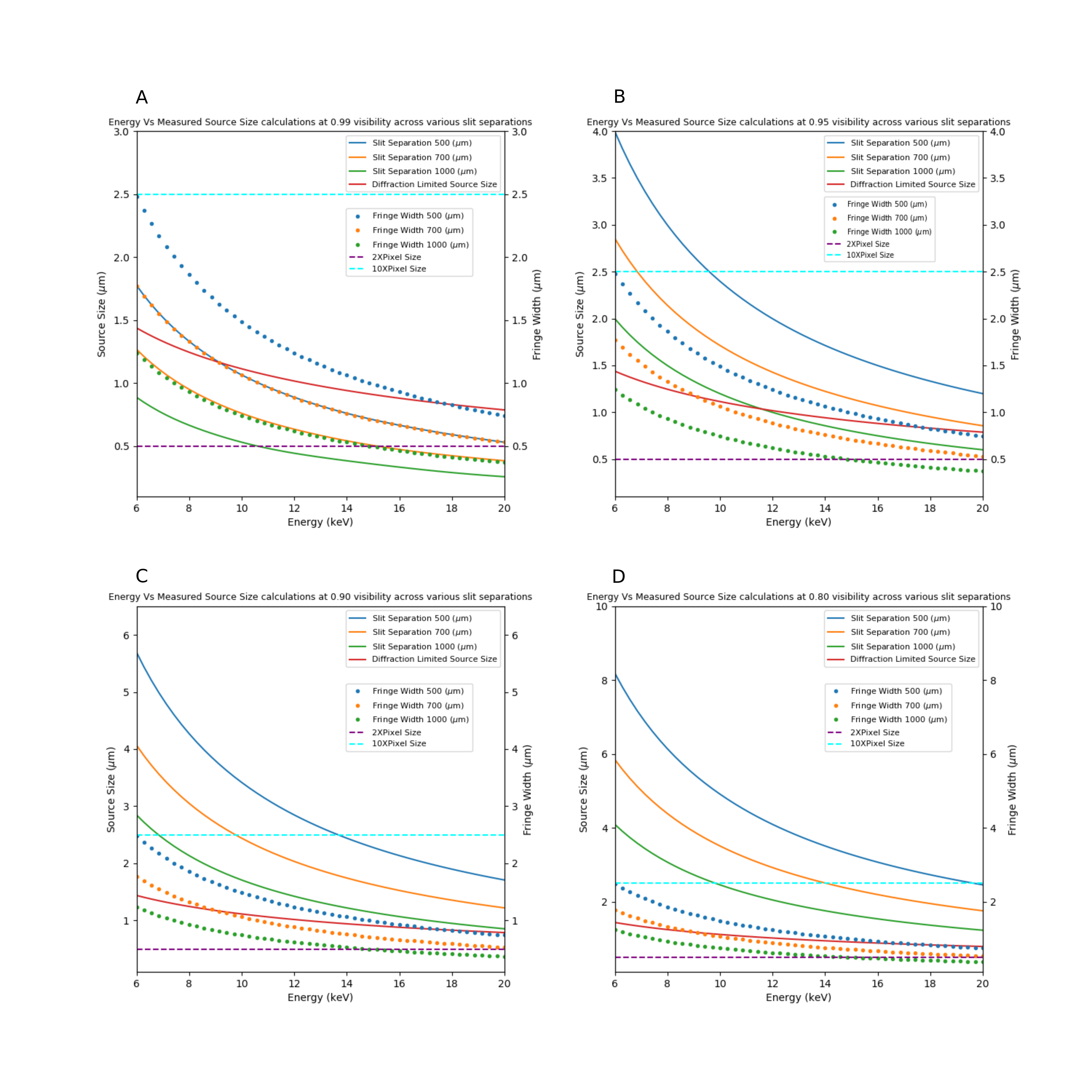}
    \caption{Theoretical calculations primarily showing the relationship between source size and energy for a given visibility value. The calculations are juxtaposed against the fringe width as a function of energy in order to provide a complete picture at a given visibility value.}
    \label{fig:discussion}
\end{figure}

Exploring the challenge of measuring very large coherence lengths or very small source sizes is just as important as investigating the limitations of measuring very low coherence lengths. State-of-the-art diffraction limited sources, in theory, boast source sizes of a few microns in the energy ranges considered in this study. The ability to measure such source sizes would definitely be the pinnacle of any coherence length measurement technique. We will put forth our theoretical calculations (Fig \ref{fig:discussion}) showing the fundamental limitations of carrying out YDS experiments to measure very small source sizes. The first limitation of such an experiment would be the inherent inability to measure source size below a point for a given visibility. In principle, the visibility can be a maximum of 1, but in practice, this is lower owing to noise. Even if we assume perfect measurements can be made, one can see (Fig \ref{fig:discussion} A) that the sampling constraints will kick in at high enough energies (1000 \si{\micro m} slit separation,$\approx$15 keV). It should be noted here that we have assumed absolute minimum detector sampling criterion (2 pixel wide, 250 nm pixel size). At 700 \si{\micro m} slit separation, the fringe width is manageable however the ability to measure diffraction limited source sizes is feasible only after certain energy limit ($\approx$9.5 keV, Fig \ref{fig:discussion} A) smaller slit separations are beyond this fundamental measurement limit. If one assumes 0.95 visibility scenario (Fig \ref{fig:discussion} B), only 1000 \si{\micro m} slit separation will be capable of measuring diffraction limited source sizes after an energy limit of $\approx$11 keV while hitting a detector sampling limit at $\approx$15 keV which is universal and visibility independent. The rest of the slit separations are beyond the fundamental measurement limit. At smaller visibility levels which are probably more reasonable, none of the slit separations explored here will be able to measure the diffraction limited source sizes. It is however important to note that these calculations were performed under an assumption that the slits to detector distance is 6m. One might expect better sampling conditions as this distance is increased usually at the expense of X-ray intensity. At this juncture it is important to make a conservative estimate on what would be the smallest source sizes that can be measured using YDS. If one assumes 0.8 as a reasonable visibility value that could be measured reliably (previously reported \cite{Takayama1998a}). While 2-pixel sampling is the absolute minimum needed sampling condition, 10-pixel sampling would ensure robustness of measurements. Taking both of these factors under consideration, one can see that 500 \si{\micro m} slit separation with 0.8 visibility to be the best candidate to obtain optimum source size measurements (Fig \ref{fig:discussion} D) and a conservative estimate of minimum measurable size would be 3-8 \si{\micro m} depending on the energy under consideration. A real-world beamline will employ multiple optical components to achieve the needed specifications of a proposed experiment. Each of these optical components will invariably introduce some amount of instabilities such as physical vibrations, ultimately manifesting into partial coherence effects. A diffraction-limited beamline might still possess effective source sizes in tens of microns far from the theoretical source size values. YDS might, therefore, satisfy the necessary condition to characterize the smallest of source sizes available as of now with the existing technology.         


\section*{Methods}
\label{section:Sec3}
\subsection*{Experimental Data Acquisition}
\label{section:Sec3.1}
\begin{figure}[h!]
    \centering
    \includegraphics[width=12cm]{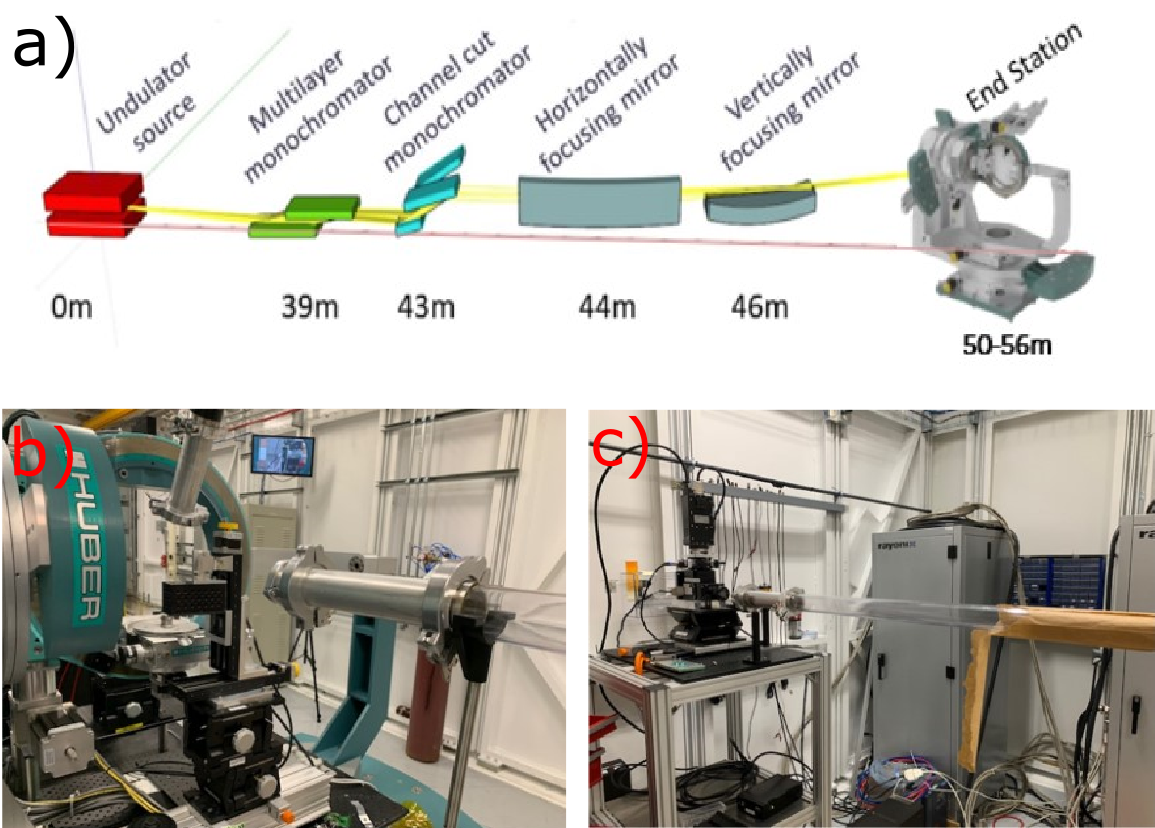}
    \caption{Displayed is the layout of the experiment performed. Subfigure (a) shows the layout of the BXDS-IVU beamline and its components. Most important to this experiment are the undulator source, the DMM, and the end station. Subfigure (b) shows the end station at the beamline where the double slit apparatus was mounted during the experiment. Subfigure (c) shows the PCO detector setup that was positioned 6m from the beamline end station. Flight tubes were employed to preserve beam intensity and reduce the effects of air scattering.}
    \label{fig:eFig1}
\end{figure}

In the current work, we employ a YDS apparatus to measure the X-ray coherence length at the BXDS-IVU beamline at the CLS. The CLS operates at an electron beam energy of \qty{2.9}{\giga\electronvolt} 
with a horizontal beam emittance of 18.1~\si{\nano \meter \cdot rad} and a vertical beam emittance of 0.1017~\si{\nano \meter \cdot rad}~\cite{Dallin2003, Cutler2007}. The beamline utilizes an in-vacuum undulator with an undulator length $b=1.58$m, deflection parameter of $K=1.79$ and offers an energy range of 5-24~\si{keV}. For this experiment, an energy range of 6.75 to 20~\si{keV} was selected ($E$ = 6.75~\si{keV}, 7.9~\si{keV}, 9.6~\si{keV}, 15~\si{keV}, 20~\si{keV}). Regarding the beamline optics, only a double-multilayer monochromator (DMM) with an energy bandpass of 0.08\% was used. The channel-cut monochromator (CCM) and Kirkpatrick-Baez focusing mirrors were bypassed to maximize the beam flux and minimize the potential degradation of coherence properties that might result from the X-ray optics instabilities (Fig.~\ref{fig:eFig1}). The slit array was fabricated using focused ion beam milling (CCEM, McMaster University) from 25~\si{\micro\meter} thick gold foil (Goodfellow Advanced Materials) and contained 2 vertical double slits with slit widths of approximately 3~\si{\micro\meter} and slit separations of 10~\si{\micro\meter} and 20~\si{\micro\meter} for the horizontal measurement, 50~\si{\micro\meter} and 100~\si{\micro\meter} for the vertical measurement. Emphasis was put on the horizontal coherence length measurements due to the larger emittance of the electron beam in the horizontal direction. For this experiment, either one of two different Pioneering in Cameras and Optoelectronics (PCO) detectors coupled with a X-ray scintillator were used with a sample-to-detector distance of 6~\si{m} as shown in Fig.~\ref{fig:eFig1}. The pixel size of this detector setup varied depending on the choice of experimental parameters. For horizontal spatial coherence length measurements, the detector used had a physical pixel size of \numproduct{9 x 9}~\si{\micro \meter}, and the vertical spatial coherence length measurements used a PCO detector with a physical pixel size of \numproduct{6.5 x 6.5}~\si{\micro \meter}. Effective pixel sizes of 3.1~\si{\micro\meter} for the horizontal coherence length measurements and 0.7~\si{\micro\meter} for the vertical coherence length measurements have been determined through measurement of the interference fringe width in the images obtained, as shown in the Supplementary Material, Section S4. Exposure times varied from 300~\si{ms} to 300~\si{s} depending on the image quality. A white beam slit (WBS), located 15~\si{\meter} upstream from the double slit array was used to constraint the source extent to dynamically change the source size when required throughout the experiment. The WBS is composed of a set of motorized horizontal and vertical tungsten slits. While the WBS is used to change the source size, it should be noted that the undulator is considered to be the GSM source for the calculation of the spatial coherence length (Equation \ref{eq:coh_len}). The double slit array was positioned 55~\si{\meter} from the undulator during the experiment (Fig.~\ref{fig:eFig1}).

\subsection*{Data Processing}
\label{section:Sec3.2}
Fringe visibility was calculated by finding the maximum and minimum values of the central most peak of the intensity curves and then applying Eq.~\ref{eq:1}. Source extension sizes were determined by fitting the experimentally observed interference intensity curves to forward modeled interference intensity curves. The forward model representing the intensity distribution of interferometric fringe blurring can be understood as a result of incoherent intensity summation of multiple coherent wavefronts originating from an extended stochastic source and can be mathematically represented as 
\begin{equation}
\label{eq:intensity}
    I(x) = I_{0}\sum_{-N}^{N} \sinc ^{2}\left(\frac{a\pi(x+N(\delta x))}{\lambda L}\right)\cos ^{2}\left(\frac{d\pi(x+N(\delta x))}{\lambda L}\right), N\in\mathbb{R},
\end{equation}
where $I_{0}$ is the incident beam intensity emitted from the center of the extended X-ray source, $a$ is the slit width, $d$ is the slit separation, and $L$ is the slit-to-detector distance. In the fitting procedure, coherent fringes are shifted in both directions by the increment $N\delta x$, here $\delta x$ is determined by the effective pixel size of the detector. This curve fitting procedure is performed iteratively over an appropriate range of $N$ such that the goodness of fit is maximized (Fig.~\ref{fig:AllFringes}, \ref{fig:FringeBlurring}). It should be noted that \textit{best fit} $N\delta x$ denotes the maximum shifted distance in the detector space, $\eta$. Source size is then calculated through $D=\eta\frac{R}{L}$. We will call this the goodness of fit method (\textbf{GOF}). Alternatively, one can also obtain the source extension sizes by generating visibility vs source size curves by sweeping through an extended range of shifted YDS intensity patterns and collecting visibility, source size ($\eta\frac{R}{L}$) values at every extension step. Once the visibility vs source size points over this extended range are obtained, one can then obtain an analytical expression between visibility and source sizes through interpolation (cubic splines). The analytical expression will enable one to obtain source sizes from experimentally obtained visibility values (Fig \ref{fig:vis_sos_intpo}). To distinguish from the previous method we will call this the interpolation method (\textbf{INTPO}). In some cases, owing to the zone of ambiguity, values obtained from the INTPO method were used as initial guesses, which were further refined by the GOF method. At times, the GOF method locks into a source size that is the ambiguous equivalent of lobe 1 but in lobe 2 or higher (Supplementary Material, Fig S11, S12, S13). Overall, we used the GOF or INTPO followed GOF protocol where accuracy was necessary, namely getting source size values from experimentally observed data, we used INTPO method in order to study big picture trends concerning visibility vs source size curves. Theoretically expected behavior between source size and visibility was obtained through equation \ref{eq:1} for a given combination of energy, slit separation and GSM to slit distance. Lastly,  we determine the absolute lower limit of the source size at a given energy as the diffraction-limited source size. Following the result from standard X-ray physics textbooks, we will define the diffraction-limited source size to be~\cite{Xray_phys_book},
\begin{equation}
\label{eq:phase}
    D_{0} = \frac{\sqrt{\lambda b}}{4\pi},
\end{equation}
where $b$ is the undulator length. Using the results of Eq.\ref{eq:intensity} and~\ref{eq:phase}, the extended source size can be determined by $D = \frac{R}{L}N(\delta x) + D_{0}$. Once the extended source size is obtained, the coherence length is determined through equation \ref{eq:coh_len} (Fig \ref{fig:CohLen}).\\

\noindent
INTPO method was used to gauze large-scale trends, the curves obtained through this method enable one to automate the detection of local maxima, minima which can then be used to determine the zone of ambiguity and necessary sampling requirements (Fig \ref{fig:zoa_samp}, \ref{fig:samp_cur}). This would further allow one to obtain the minimum source size before ambiguity (\textbf{MSBA}) and maximum visibility before ambiguity (\textbf{MVBA}) points in an automated fashion which allowed us to generate large range MSBA, MVBA behavior  for both visibility and form categories over varying energy, slit width with fine granularity (Fig \ref{fig:amb_cur}).\\  

\noindent
YDS experiment feasibility studies shown in Fig \ref{fig:discussion} were obtained through equation \ref{eq:1}. The theoretical visibility vs source size curves were generated from a wide enough range of source sizes, which were then taken through the interpolation method (cubic splines) to find a spline function that gives source size as a function of visibility values at a given energy. We then repeated this process at different slit separations, energy values collecting source sizes for a given visibility at every step.

\bibliography{sample}

\section*{Acknowledgements}
The research described in this article was carried out at the Canadian Light Source, a national research facility of the University of Saskatchewan, which is supported by the Canada Foundation for Innovation (CFI), the Natural Sciences and Engineering Research Council (NSERC), the National Research Council (NRC), the Canadian Institutes of Health Research (CIHR), the Government of Saskatchewan, and the University of Saskatchewan. We gratefully acknowledge the support and help of Prof. Gianluigi Botton of McMaster University in realizing this experimental activity. We gratefully acknowledge Dr. Drew Bertwistle for his help with organizing this beam time. We gratefully acknowledge Dr. Sergey Gasilov and Dr. Adam Webb for the use of their detectors and their assistance in the experiment. We gratefully acknowledge Burke Barlow for collecting the SEM images for our analysis.

\section*{Author contributions statement}
VSCK conceptualized the project and organized the experiments. RC, VSCK, NA collected data during experiments. RC and VSCK performed the data analysis. RC and VSCK wrote the manuscript. All authors contributed to the review and editing of the manuscript.

\section*{Additional information}
The authors declare no competing interests. The online version of this article contains supplementary information that can be found at [TBD]. Correspondence and requests for materials should be addressed to the primary or corresponding authors.


\newpage
\newpage

\newpage

\newpage

\newpage

\newpage

\newpage

\newpage

\newpage

\newpage

\newpage
\includepdf[pages=-]{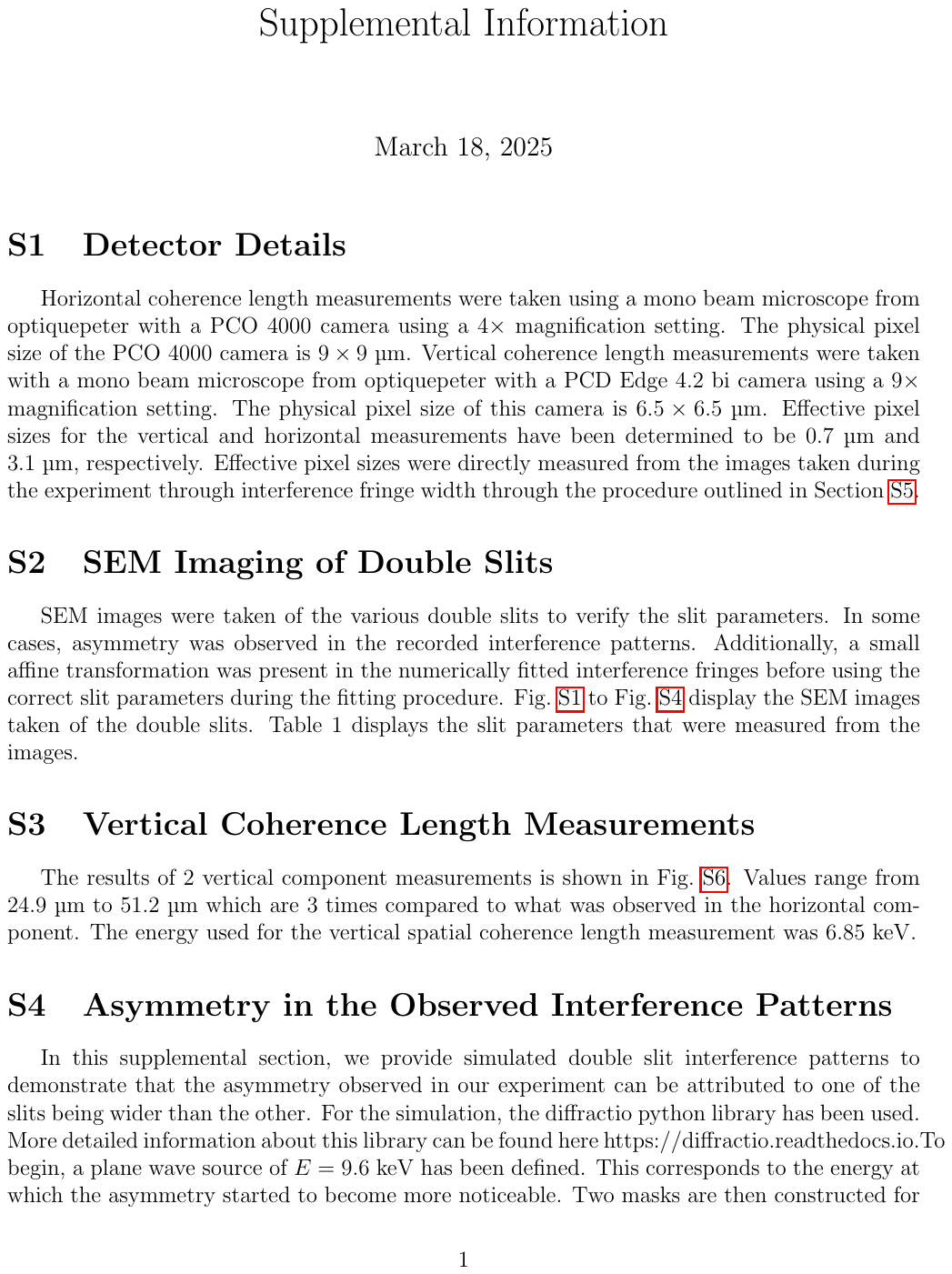}


\end{document}